\documentclass{cda}
\usepackage{array}

\begin{document}

\title{Uma breve reflexão sobre a Mostra de Astronomia do Espírito Santo}

\author{Adriano M. Oliveira$^1$}\email{adriano.oliveira@ifes.edu.br}
\author{Lúcia Horta$^{2}$}\email{lucia-horta@hotmail.com}

\instituto{$^1$Instituto Federal do Esp\'irito Santo (Ifes), Guarapari - ES, Brasil}
\instituto{$^2$Secretaria Estadual de Educação do Espírito Santo (SEDU-ES), Guarapari - ES, Brasil}

\abstract{In this paper we take a brief discussion about the two editions of State Show of Astronomy and its influence in process of student formation. One comparison between numbers of projects registrants and also total teachers registrants, of these editions, we see that is necessary to create forms for the teacher of public school could dedicate time for orientation activities. Once these practices help to better the teaching-learning relationship.}


\resumo{Esse artigo traz um breve relato sobre as duas edições da Mostra Estadual de Astronomia e levanta uma discussão sobre a influência que esse tipo de evento tem no processo de formação discentes. 
As comparações entre as quantidades tanto de trabalhos inscritos quanto de professores orientadores, participantes das duas etapas, revelam a necessidade de criar meios para que o corpo docente da rede pública de ensino possa dedicar tempo a atividades de orientação.
Visto que, essas práticas contribuem para o melhoramento da relação ensino-aprendizagem.}

\pchave{Mostra, Ensino Médio, Ensino, Pesquisa, Extensão}

\maketitle

\oautor{Oliveira, A. M. et al}
\section{INTRODU\c C\~AO}

O movimento de realização de feiras científicas teve início por volta dos anos 1960 \cite{mancuso2006feiras}.
Esse evento é caracterizado, basicamente, pela exposição de trabalhos técnico-científicos elaborados por estudantes sob a orientação de professores. 
Gonçalves \cite{gonccalves1989feiras} diz que:
\begin{quote}
as feiras de ciências consistem na apresentação de trabalhos e na relação expositor-visitante na qual são apresentados materiais, objetivos, metodologia utilizada, resultado e conclusões obtidas.
No caso da mostra, temos, palestras e um acompanhamento dos melhores trabalhos apresentados nesses eventos. 
Os trabalhos são elaborados de forma a serem  planejados e executados pelos estudantes com o professor como orientador do processo.
\end{quote}

Dentro desse contexto a mostra é um evento mais completo que a feira, já que, além da exposição, também reserva espaços para discussões e palestras, por exemplo. 
Em vista disso, a mostra científica se torna um complemento dos trabalhos realizados nas feiras e, por esse motivo, escolhemos estudar as melhorias que a implementação de uma mostra estadual, a Mostra de Astronomia do Espírito Santo (MAES), pode trazer para os alunos do ensino médio, em particular os da rede pública de ensino, sendo este o primeiro artigo comparativo entre as duas primeiras edições do evento.

Durante o processo de preparação, estudo e montagem dos trabalhos, a dinâmica tradicional de ensino é mudada, enquanto os professores deixam o papel de transmissor do conhecimento, passando a atuar como mediadores, os alunos se tornam protagonistas da relação ensino-aprendizagem, propondo hipóteses e aprendendo com seus erros.
Essa forma de atuação está alinhada com a metodologia de ensino por investigação que, por sua vez, é uma alternativa para o método tradicional de ensino, permitindo uma aproximação dos conteúdos ministrados em sala de aula às habilidades previstas nas Diretrizes Curriculares Nacionais (DCN) da educação básica \cite{dediretrizes}.
Além disso, trabalhando com esse método, a pesquisa científica direcionará tanto os processos de confecção dos trabalhos para a mostra quanto os conteúdos curriculares previstos na legislação da educação básica.

Dessa forma, a participação em mostras científicas pode promover mudanças significativas nos alunos, uma discussão mais aprofundada sobre esse assunto pode ser encontrada em  \cite{mancuso1993evoluccao,lima2008feiras}. 
A apresentação dos trabalhos é conduzida pelos alunos, colocando-os na posição de protagonistas e, ainda, levam a toda comunidade, que visita a mostra, todo o conhecimento produzido ao longo da preparação dos pôsteres e seminários.
Mostrando que a utilização da mediação pode ensinar e de transmitir conhecimento, além de: modificar a tradicionalidade da sala de aula formal e, também, atingir mais pessoas com mais efetividade.
Ainda, com a implementações de ações como esta, atende-se a recomendação das DCN no tocante ao desenvolvimento da capacidade de pesquisa como parte importante na `` busca da (re) construção do conhecimento'' \cite{dediretrizes}.
Além disso, amplia a atuação do professor, segundo \cite{demo2011educar}:
\begin{quote}
    educar pela pesquisa tem como condição essencial primeira, que o profissional da educação seja pesquisador, ou seja maneje a pesquisa como princípio científico e educativo e a tenha como atitude cotidiana.
\end{quote} 
Ou seja, é necessário que o professor oriente o aluno para que sua pesquisa tenha uma direção e produza um trabalho rico em conhecimento.
Nesse caminho, as Mostras Científicas agem como disseminadoras dos trabalhos e do conhecimento produzido, para públicos diversos.
Devido a relevância desses eventos, tanto para a formação dos alunos quanto para a mudança da prática docente, vários trabalhos têm usado eles como base de estudo para suas pesquisas, por exemplo, o levantamento bibliográfico, entre os anos 2008-2018, de caráter exploratório qualitativo realizado sobre feiras de ciências \cite{dafeiras}, a discussão sobre a importância do protagonismo do aluno no
contexto escolar a partir de sua participação em mostras com temáticas de astronomia \cite{bernardesmostras} e  uma revisão bibliográfica sobre o ensino de ciências e as feiras, buscando ressignificar o aprendizado significativo e o ensino mais dialógico \cite{bertoldo2016feiras}, ou seja, a relação ensino-aprendizagem mais fundamentada no diálogo entre as partes (professor e aluno) envolvidas nesse processo.

Diante disso, faremos aqui um breve relato acerca da influência que as feiras e mostras científicas têm na efetivação de futuros projetos de pesquisa desenvolvidos no Espírito Santo, em particular analisaremos aqueles trabalhos que participaram da MAES. 
O evento estadual é realizado anualmente e está em sua terceira edição. 
O objetivo do estudo foi fazer um acompanhamento dos trabalhos apresentados nas duas edições do evento (2018-2019) traçando: mudanças, adaptações necessárias, apresentando resultados e possibilidades para novas pesquisas.

\section{MAES 2018}
\begin{figure}[h!]
\centering
\includegraphics[scale=0.3]{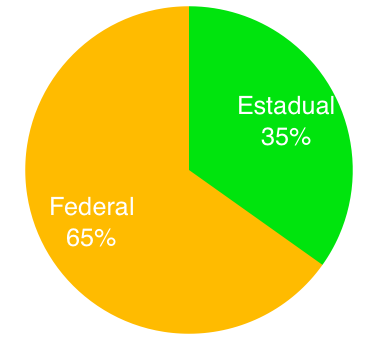}
\caption{A MAES2018 teve participação de escolas federais e estaduais. A fatia verde, correspondente a 35\%, refere-se a porcentagem dos trabalhos produzidos por alunos da rede estadual e a parte em amarelo está ligada às propostas do Ifes, cerca de 65\% dos projetos foram realizados por alunos ligados a esta instituição de ensino.}
\label{fig:FotoDivEsc2018}
\end{figure}

Na edição 2018 da Mostra de Astronomia do Espírito Santo (MAES), primeiro ano do evento, 89 (oitenta e nove) trabalhos foram inscritos. 
Destes, 65,2\% decorrentes de Institutos Federais de ensino (Ifes) e o restante, 34,8\%, tiveram como autores alunos das escolas estaduais, como mostrado na figura (\ref{fig:FotoDivEsc2018}).
Não tivemos projetos de escolas particulares nem de alunos do 9$^o$ ano, na referida edição.
Logo, todos os projetos foram elaboradas e apresentados por alunos que estavam cursando o ensino médio durante o ano de 2018, sendo os três melhores, após duas fases, premiados com bolsas CNPq de PIBIC-Jr (ICJ), para o ano de 2019.  

\begin{figure}[h!]
\centering
\includegraphics[scale=0.3]{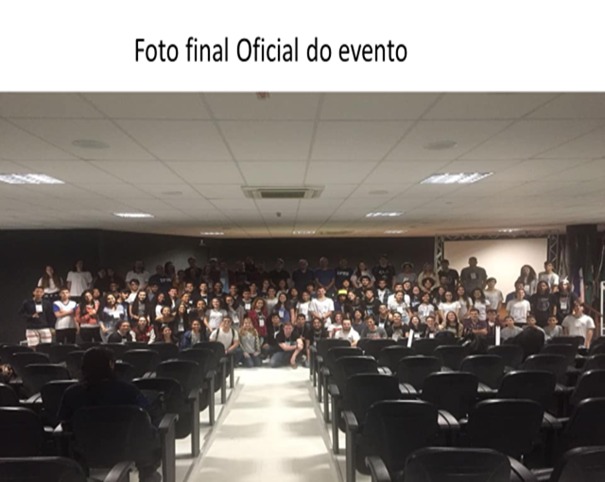}
\caption{Essa é a foto oficial da edição 2018 da Mostra de Astronomia do Espírito Santo (MAES2018).}
\label{fig:FotoOficial2018}
\end{figure}

A primeira fase foi dividida em três etapas, contudo os trabalhos foram inscritos apenas para a etapa de Cariacica, onde 32 (trinta e dois) professores e cerca de 350 (trezentos e cinquenta) alunos, participaram do evento, a foto oficial desta etapa é mostrada na figura (\ref{fig:FotoOficial2018}).
Nas etapas interioranas, realizamos uma bateria de seminários ministrados pelos professores convidados, observação do céu noturno com telescópios, um {\it{tour}} pelo mapa de Marte, oficina de foguetes e de caça a meteoros. 
Toda a estrutura do evento foi coloca à disposição do público.

Desta primeira fase foram selecionados 15 (quinze) trabalhos para a fase final, onde os projetos foram apresentados de forma oral a uma banca de especialistas com destacada relevância na comunidade científica, que atuaram como avaliadores.
A segunda fase, teve duração de dois dias e ocorreu um mês depois da seletiva, além dos seminários, grande parte da estrutura acima citada foi novamente montada para atender escolas e toda a comunidade interessada.
Ao final das apresentações, 15 (quinze) alunos foram agraciados com bolsas de ICJ e tiveram seus trabalhos acompanhados pelos professores e pesquisadores ligados ao Cosmo-Ufes. 
Além disso, participaram de encontros mensais, durante o ano de 2019, que ocorreram tanto no Observatório Astronômico do Ifes Guarapari (OAIG) quanto em parques estaduais, centros históricos - o que lhes permitiu: melhorar seus projetos, ampliar o conhecimento sobre a cultura, fauna, flora capixaba, e melhorar a qualidade do conhecimento adquirido.
 Ademais, esse grupo de alunos participou de palestras envolvendo tópicos de astrofísica, astronomia, astronáutica, cosmologia e foram convidados a participar de todas as ações do Cosmo-Ufes, tais como o Verão Quântico, o Inverno Astrofísico e a comemoração dos 100 anos do Eclípse de Sobral, todos ocorreram durante a vigência da bolsa ICJ. 
 Por fim, participaram das ações abertas ao público de observações do céu e receberam orientação para futuros estudos de elementos teóricos ligados ao tema da mostra.
 Ou seja, os bolsistas foram imersos tanto nos congressos científicos como na condução de ações com a comunidade externa.

\section{MAES 2019}
\begin{figure}[h!]
\centering
\includegraphics[scale=0.3]{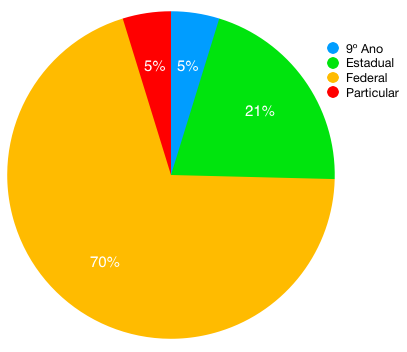}
\caption{Esse gráfico mostra como foi a divisão das escolas participantes da MAES2019. Em vermelho e azul, com 5\% do total de trabalhos, estão as escolas particulares e escolas municipais (alunos de 9$^o$ ano), respectivamente, ambas com três trabalhos apresentados. Em verde, correspondendo a 21\% do total de trabalhos, está representado os trabalhos da rede estadual, com 13 (treze) trabalhos apresentados, e em amarelo os trabalhos do Ifes, cerca de 70\% dos trabalhos, ou 44 (quarenta e quatro) projetos, foram confeccionados por alunos ligados a esta instituição de ensino.}
\label{fig:FotoDivEsc2019}
\end{figure}

\begin{figure}[h!]
\centering
\includegraphics[scale=0.3]{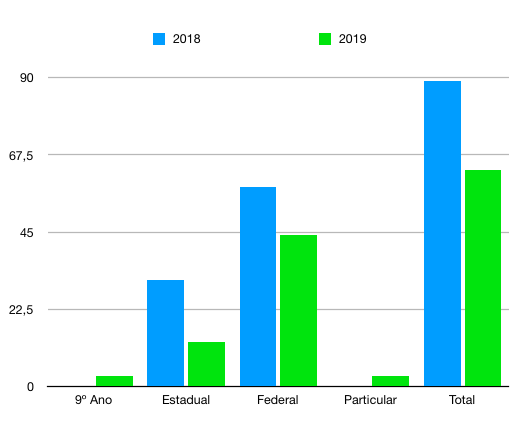}
\caption{Esse gráfico mostra um comparativo entre a quantidade de trabalhos apresentados nas duas edições da MAES, por rede de ensino, em azul estão as quantidades referentes ao ano de 2018 e em verde os dados do ano de 2019.}
\label{fig:Comp1819}
\end{figure}

Para a edição 2019, algumas mudanças foram implementadas, por exemplo, os melhores trabalhos não foram premiados com bolsas, mas sim com medalhas e troféu, já que não conseguimos recursos para mantê-las nas agências de fomento daquele ano.
Por outro lado, dividimos a competição em categorias e ampliamos o público alvo, permitindo a participação tanto de alunos do 9$^o$ ano do ensino fundamental quanto de escolas privadas.
Dessa forma, ampliamos a diversidade de escolas participantes, como mostra a figura (\ref{fig:FotoDivEsc2019}), tivemos 5\% dos trabalhos confeccionados pelas escolas privadas e a mesma quantidade por escolas municipais (alunos de 9$^o$ ano), já os projetos das escolas estaduais corresponderam a 21\% do total e o Ifes inscreveu 70\% dos trabalhos. 
Portanto, tivemos representantes de todo o público alvo, do mesmo modo que na edição 2018. 
Analisando os números apresentados nas figuras (\ref{fig:FotoDivEsc2018}) e (\ref{fig:FotoDivEsc2019}), verificamos que o Ifes ampliou sua participação em quase 5\%, enquanto as escolas estaduais tiveram uma redução de 14\%.
Já o número total de inscritos sofreu redução de 28,1\%, saindo de 83 (oitenta e três) inscritos em 2018 para 64 (sessenta e quatro) em 2019.
Uma comparação entre as quantidades de trabalhos por rede de ensino é apresentada na figura figura (\ref{fig:Comp1819}).
Do mesmo modo, o número de professores orientadores também reduziu, saindo de 32 (trinta e dois) e passando para 14 (quatorze), o que corresponde a uma queda de 56,25\%.

\begin{figure}[h!]
\centering
\includegraphics[scale=0.2]{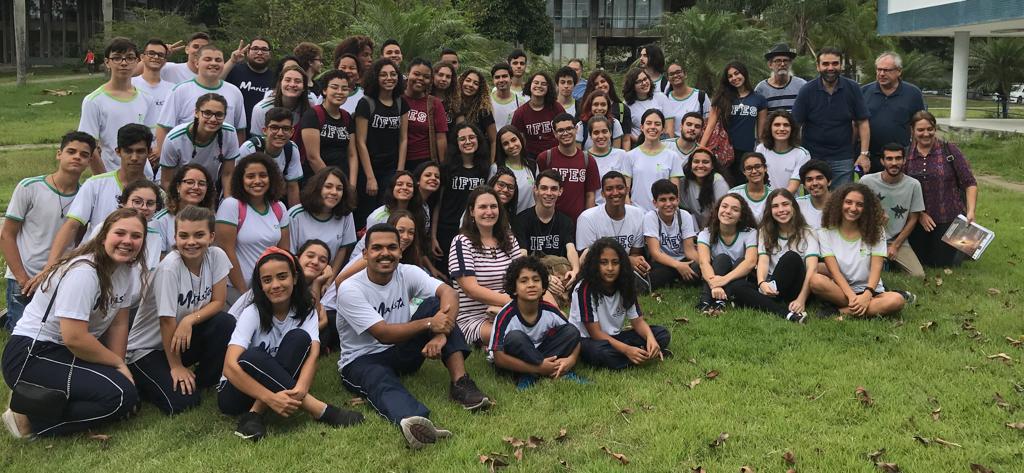}
\caption{Essa é a foto oficial da edição 2019 da Mostra de Astronomia do Espírito Santo (MAES2019).}
\label{fig:FotoOficial2019}
\end{figure}

Apesar dessa queda, tivemos a percepção de que os trabalhos apresentados durante a MAES2019 tiveram um acréscimo de complexidade, quando comparadas as duas edições do evento, mostrando uma melhora na relação ensino-aprendizagem e na mediação feita pelos professores orientadores.
Além do mais, cerca de 30\% das inscrições vieram das escolas do interior do estado. 
Mesmo em menor número esses projetos mostram certo avanço tanto na rede de comunicação criada no interior do estado quanto no envolvimento das escolas para participar da MAES. 
\begin{figure}[h!]
\centering
\includegraphics[scale=0.3]{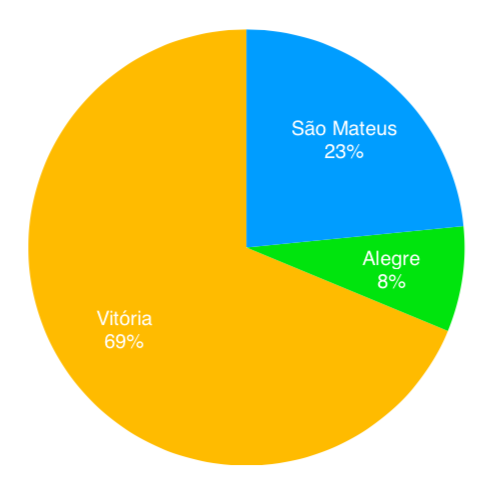}
\caption{Nessa figura está representada a distribuição dos projetos cadastrados para as etapas de Alegre (com 5\%), São Mateus (com 23\%) e Vitória (com 69\%), para a primeira fase da Mostra de Astronomia do Espírito Santo 2019 (MAES2019).}
\label{fig:Etapas2019}
\end{figure}
Outra mudança gerada pela limitação de recursos foi a redução na quantidade de atividades ofertadas ao público.
Para essa edição mantivemos as apresentações dos trabalhos, seminários dos professores visitantes (que também atuaram como avaliadores dos projetos nas duas fases) e a observação do céu noturno com telescópios.
Uma foto da etapa de Vitória do evento é apresentada na  figura (\ref{fig:FotoOficial2019}), esse foi o local onde tivemos a maior concentração de trabalhos durante a primeira fase da Mostra.

Desta primeira fase, foram selecionados 21 (vinte e um) trabalhos para a fase final, sendo 13 (dezesseis) do Ifes, 5 (seis) da rede estadual, 1 (um) da rede municipal  e 2 (dois) da rede privada, que ocorreu no Observatório Astronômico do Ifes Guarapari e teve a duração de dois dias.
Nesta fase, os projetos foram apresentados de forma oral para uma banca de avaliadores, pesquisadores ligados ao Cosmo-Ufes, que também ministraram seminários sobre temas que abordaram assuntos de ``Buracos Negros: Fotografando o invisível'' e ``Explicando o Nobel de física 2019'', durante os dois dias de evento. Após essa etapa os destaques foram:
\begin{itemize}
    \item {\bf{Escola Particular (Medalha de Ouro)} - } TSI -- Telescópios para a síndrome de Irlen, produzido por alunos do Marista.
    Essa síndrome causa um desequilíbrio na capacidade de adaptação à luz e está ligada ao deficit na leitura, o que também causa problemas no uso de telescópios convencionais. 
    Nesse sentido, o projeto foi fundamentado na adaptação de um telescópio para que portadores desta síndrome tivessem a oportunidade de enxergar as estrelas sem os prejuízos desta. Para isso foi criada uma lente com características semelhantes aos óculos usados por pessoas que possuem essa sensibilidade.

    \item {\bf{Escola Estadual (Medalha de Ouro)} - } Registro de meteoros na constelação de Hydra. Esse trabalho apresentou um estudo referente a captação do registro dos meteoros vindos da região da constelação de hydra, realizada pela estação de videomonitoramento de pequenos objetos da escola Dr. Silva Mello. Com as imagens o aluno encontrou os padrões da quantidade de radiantes registrado mensalmente, da velocidade média, magnitude e direção.

    \item {\bf{Escola Municipal (Medalha de Ouro)} -} Celebridades e descobertas astronômicas, produzido por alunos do 9 ano. 
    O Projeto buscou realizar uma viagem no tempo e apresentou os grandes nomes da Astronomia, suas contribuições e descobertas. Além disso, explorou os saberes adquiridos pelos alunos sobre conceitos, invenções e história da astronomia, bem como a evolução do conhecimento sobre essa área.
    O recorte escolhido foi a compreensão do sistema solar e fenômenos que acontecem no nosso dia-a-dia.
    Buscou-se incentivar tanto a criatividade quanto um alinhamento às habilidades e competências previstas na BNCC, além da prática experimental que complementa o conteúdo teórico e leva a interdisciplinaridade, via investigação.

    \item {\bf{Escola Federal (Medalha de Ouro e Melhor trabalho da MAES2019)} -} Astrologia não é uma ciência, produzido por alunos do Ifes, campus Vitória. O trabalho trouxe a luz uma discussão acerca da influência das estrelas sobre a vida na Terra, buscando desmistificar conceitos populares fundamentando sua pesquisa em dados científicos.
    
    \item {\bf{Escola Federal (Medalha de Prata} -} Física e Astronomia, que abordou a física dos telescópios, da teoria à construção de um telescópio newtoniano. O trabalho foi produzido por alunos do Ifes, campus Guarapari.
    
     \item {\bf{Escola Federal (Medalha de Bronze} -}  A morte térmica do Universo, outro trabalho do Ifes, campus Guarapari.
\end{itemize}

\section{CONSIDERA\c C\~OES FINAIS}

A mostra é uma importante aliada para as mudanças metodológicas tradicionais.
Elas Incentivam o uso de metodologias investigativas como uma alternativa para melhorar a qualidade da relação ensino-aprendizagem e retorna ao aluno a responsabilidade por ampliar seu próprio conhecimento.
A Mostra de Astronomia do Espírito Santo, cuja terceira edição ocorrerá durante o ano de 2020, é um exemplo desse tipo de ação que exercita o protagonismo estudantil.

A partir de Uma comparação entre os trabalhos apresentados durante as duas edições do evento, percebemos que houve uma melhora significativa na qualidade das propostas e da organização do evento. 
Ainda, percebemos um pequeno avanço, oriundo das mudanças na forma de divulgar o evento, já que tivemos escolas do interior do estado participando da segunda edição do evento, fato que não ocorreu durante o ano de 2018.
Contudo, tivemos uma redução de quase 30\% no número de trabalhos inscritos e de mais de 50\% de professores orientadores, com maior impacto na participação das escolas estaduais, o que de certo modo pode ter sido influenciado pela não ofertada de bolsas para os melhores trabalhos no ano de 2019.
Mas, para um evento estadual, julgamos baixo o número de inscritos, o que deixa clara duas fragilidades: (1) a divulgação da MAES ainda não é a ideal e (2) a falta de interesse dos professores do ensino básico em participar desse tipo de eventos, fato comprovado pelo baixo número de professores orientadores durante a última edição do evento, apenas 14 (quatorze). 

Sendo assim, julgamos que é necessário criar meios para que os docentes possam se sentir atraídos em participar desse tipo de evento, em particular os da rede estadual. 
Uma possibilidade seria a premiação dos professores e, de algum modo, permitir a alocação de carga-horária, talvez nos moldes de planejamento, para orientação de equipes de alunos interessadas em participar de mostras científicas.
Esse tipo de incentivo, juntamente com ações de formação continuada docente, que fomentem o pensar científico atrelado ao ensino de astronomia utilizando metodologias ativas, como a proposta no trabalho \cite{oliveira2020}, podem ajudar a alinhar os conteúdos de astronomia às habilidades previstas na BNCC e na DCN, o que além de ampliar o número de trabalhos inscritos na MAES e similares, vai melhorar significativamente a qualidade da relação ensino-aprendizagem.

\section{AGRADECIMENTOS}
Os autores agradecem aos pesquisadores que, de algum modo, colaboraram  para a realização dessas duas primeiras edições do evento, de modo especial, ao digníssimo Professor Doutor Emérito Antônio Brasil que nos presenteou com sua ilustre presença durante a MAES 2019, com muito humor e sapiência invejável. 
Certamente, um dos maiores físicos da nossa época que nos deixou uma lição de humildade e amor pela ciência. 
Infelizmente não poderemos mais ter o prazer de sua presença nas próximas edições da Mostra de Astronomia do Espírito Santo, mas levaremos a cada edição os ensinamentos por ele deixado.
Em tempo, agradecemos a Capes, Fapes, Ufes e Ifes pelo apoio e suporte dado durante o evento.

\end{document}